\documentclass[conference]{IEEEtran}
\IEEEoverridecommandlockouts

\usepackage{blindtext}

\usepackage{etoolbox}
\usepackage{lipsum} %<---- For dummy text

\usepackage{siunitx}
\usepackage{pifont}
\usepackage{graphicx}
\usepackage{amsmath, amssymb,textcomp}
\usepackage{algorithm} 
\usepackage{booktabs}
\usepackage{url}
\usepackage{subcaption} 
\usepackage{algpseudocode}
\usepackage{paralist}
\usepackage{color}
\usepackage{relsize}
\usepackage{stmaryrd}
\usepackage{epsfig} % for postscript graphics files
\usepackage{bm} % assumes amsmath package installed
\usepackage{listings,lstautogobble}
\usepackage{color}
\usepackage{mathtools}
\usepackage{mathrsfs}
\usepackage{xspace}
\usepackage{soul,xcolor}
\usepackage{float}
\usepackage{verbatim}
\usepackage{epstopdf}
\usepackage{listings}
\usepackage{parcolumns}
\usepackage{color}
\usepackage{mathtools}
\usepackage{mathrsfs}
\usepackage{xspace}
\usepackage{soul,xcolor}
\usepackage{float}
\usepackage{verbatim}
\usepackage[utf8]{inputenc}
\usepackage{calrsfs}
\usepackage{rotating,multirow}
\usepackage{caption}

\usepackage[most]{tcolorbox}
\usepackage{inconsolata}
\usepackage{float} % here for H placement parameter
\DeclareMathOperator*{\argminB}{argmin}   % Jan Hlavacek

\newtcblisting[auto counter]{samplelisting}[2][]{sharp corners, 
    fonttitle=\bfseries, colframe=gray, listing only,
    listing options={basicstyle=\fontsize{7.3}{12}\ttfamily,xleftmargin=0em, aboveskip=0em, belowcaptionskip=0em,,
    belowskip=0em,showstringspaces=false,
    breaklines=false,language=java},   
    title=Function \thetcbcounter: #2, #1}

\newcommand{\sys}{SecSens} 
\newcommand{\algone}{SecEKF} 
\newcommand{\algtwo}{SecOPT} 
\newcommand{\origekf}{OrigEKF}

\begin{document}
%\title{S-SLATS: Secure Simultaneous Localization and Time Synchronization}
\title{\sys{}: Secure State Estimation with Application to Localization and Time Synchronization}

%0.9cm] 
\author{\IEEEauthorblockN{Amr Alanwar}
\IEEEauthorblockA{\textit{University of California, Los Angeles} \\
alanwar@ucla.edu}\\
\IEEEauthorblockN{Jo\~{a}o Hespanha}
\IEEEauthorblockA{\textit{University of California, Santa
Barbara} \\
%, Country \\
hespanha@ece.ucsb.edu}\\
\and
\IEEEauthorblockN{Bernhard Etzlinger}
\IEEEauthorblockA{\textit{Johannes Kepler University} \\
%Linz, Austria \\
bernhard.etzlinger@jku.at}\\
\IEEEauthorblockN{Mani Srivastava}
\IEEEauthorblockA{\textit{University of California, Los Angeles} \\
%Santa Barbara, Country \\
mbs@ucla.edu}
\and
\IEEEauthorblockN{Henrique Ferraz}
\IEEEauthorblockA{\textit{University of California, Santa
Barbara} \\
henrique@ece.ucsb.edu}
}
%  \vspace{-5cm}
%\thanks{test}

% I used the etoolbox package to replace the default 0.5\baselineskip with -2\baselineskip
\makeatletter
\patchcmd{\@maketitle}
  {\addvspace{0.5\baselineskip}\egroup}
  {\addvspace{-2\baselineskip}\egroup}
  {}
  {}
\makeatother

\maketitle

%\section*{Abstract}
\begin{abstract}

Research evidence in Cyber-Physical Systems (CPS) shows that the introduced tight coupling of information technology with physical sensing and actuation leads to more vulnerability and security weaknesses. But, the traditional security protection mechanisms of CPS focus on data encryption while neglecting the sensors which are vulnerable to attacks in the physical domain. Accordingly, researchers attach utmost importance to the problem of state estimation in the presence of sensor attacks. In this work, we present \sys{}, a novel approach for secure nonlinear state estimation in the presence of modeling and measurement noise.
%, which is applied to geographically distributed sensors in the presence of attacks, modeling noise, and measurement noise. \sys{} 
\sys{}
consists of two independent algorithms, namely, \algone{} and \algtwo{}, which are based on Extended Kalman Filter and Maximum Likelihood Estimation, respectively. We adopt a holistic approach to introduce security awareness among state estimation algorithms without requiring specialized hardware, or cryptographic techniques. We apply \sys{} to securely localize and time synchronize networked mobile devices. \sys{} provides good performance at run-time several order of magnitude faster than the state of art solutions under the presence of powerful attacks. Our algorithms are evaluated on a testbed with static nodes and a mobile quadrotor, all equipped with commercial ultra-wide band wireless devices.

\end{abstract}


%We are able to 

%Therefore, we propose D-SLATS, a framework comprised of three different and independent algorithms to
%jointly solve time synchronization and localization problems in a
%distributed fashion.  two algorithms are based mainly on
%the distributed Extended Kalman Filter (EKF) whereas the third one
%uses optimization techniques. No fusion center is required, and
%the devices only communicate with their neighbors. \let\thefootnote\relax\footnote{$^\bullet$Equal contribution and ordered alphabetically}
%The proposed methods are evaluated on custom Ultra-Wideband communication Testbed and a quadrotor, Our algorithms achieve up to three microseconds time synchronization accuracy and 30 cm localization error
\section{Introduction}

Last decades have witnessed a proliferation in Cyber-Physical Systems (CPS) for observing a variety of physical phenomena and provide appreciate control actions. Consider, for instance, a wireless sensor network which offers many advantages and services in emergency rescue, homeland security, military operations, habitat monitoring, and home automation services. Such a network typically consists of spatially distributed sensor nodes to monitor the state of network \cite{conf:wirelessnetwork}. Furthermore, modern vehicles are another example for CPS which present a heterogeneous system for sensing dynamics and providing the required control actions \cite{conf:car}. 
The new trend in vehicle design is to move from typical isolated control systems to more open automotive architectures that would introduce new services such as remote diagnostics and code updates, and vehicle-to-vehicle communication in order to improved quality and reliability of such systems \cite{conf:autosar}.

To use these systems securely and efficiently, we need a paradigm shift in algorithmic design. For instance, focusing only on the use of data encryption is the wrong approach to assure CPS security. An interdisciplinary approach is needed that brings a diverse set of disciplines to bear on the secure design process of CPS algorithms. Such an approach for CPS should take care of security vulnerabilities that are easily exploitable in life-critical applications and lead to catastrophic consequences. These security vulnerabilities target sensing software, hardware, and physical signals due to the new interaction between information technology and physical world \cite{conf:moreattacks}. Such security vulnerabilities result from the dramatic increase in the set of sensors functionalities and system design complexity.
\begin{figure}[t]
\centering
\includegraphics[width = 0.5\textwidth]{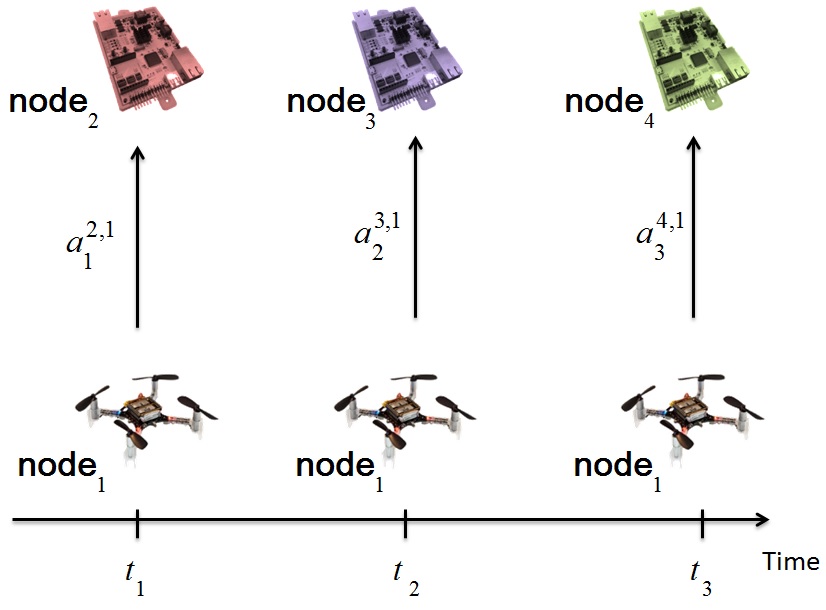}
\caption{Time dispersion of measurements. At time $t_i$ of step $i$ a measurement is taken on one link in the network, where all the links are under time varying attacks.}
\label{fig:toy}
%  \vspace{-0.4cm}
\end{figure}
\vspace{-0.1cm}

Attacks launched in the cyber domain led to calamitous consensuses during the past decades \cite{conf:cyberattacktypes}. For instance, one popular attack is the StuxNet attack on Supervisory Control and Data Acquisition (SCADA) controllers system which is used in industrial processes control \cite{conf:stuxnet1,conf:stuxnet2}. Other security issues on SCADA  networks are shown in \cite{conf:SCADAsecurity}. Examples of wireless network attacks are shown in \cite{conf:wirelessattacks}. Furthermore, the fragility of the underlying system structure of today's vehicles are exploited to a reasonable extent in \cite{conf:automotive1,conf:automotive2}, which target Electronic Control Units (ECU). They showed that today's vehicles are not resilient  against attacks mounted on its internal components. Thus, an attacker can easily disrupt the operation of critical functions of a vehicle. This problem is, even more,  emerging with the rise of self-driving cars \cite{conf:automotive3}. Also, attacks on analog sensors which have increasingly become an indispensable part of many modern systems  are shown in \cite{conf:ghost}. Similarly, attacks on CPS could affect the underlying infrastructure as shown in the Maroochy Water breach \cite{conf:maroochywaterbreach}. Moreover, the vulnerability of drones is a very dangerous threat if attackers can take control of these deadly weapons \cite{conf:drones, conf:dronesattack}. Also, consider the case of generating a fake GPS signal that appears identical to those sent out by the real GPS \cite{conf:spoofers} to take over some unmanned aircraft \cite{conf:uavgpsattack}. 
%\blfootnote{$^*$Bernhard Etzlinger and Henrique Ferraz are equally contributing authors and ordered alphabetically.}
%For instance, the attacker has infected the cockpits of America’s Predator and Reaper drones \cite{conf:dronesattack}. %% BE: citations added to the previous sentence
% We refer the reader to \cite{conf:cyberattacktypes} for more details and attacks. %% BE: citation moved to first sentence of paragraph

%Therefore, depending only on the cyber-security techniques to protect CPS can lead to catastrophic failures.

While considerable research has explored CPS security
using cryptographical techniques, it is not sufficient to ensure for securing CPS. Such cyber-security techniques cannot protect against compromised physical environment around a sensor node, which may inject a corrupted signal. Therefore, there is an emerging need for secure systems \cite{conf:needforsecurity} while considering new techniques. Thus, researchers came up with techniques that address the problem of \textbf{secure state estimation} and \textbf{intrusion detection} under the sensor, actuators, and communication network attacks. Secure state estimation enables to estimate the state of the CPS from corrupted/attacked measurements, while intrusion detection aims to detect the presence of sensor attacks.
%which can be called \textbf{intrusion detection}.

CPS offer critical services which would require meeting stringent timing and maintaining some guarantees on the accuracy of the state estimation during operating hours. However, CPS have power consumption constraints \cite{conf:powerconstraint}, limitations in terms of bandwidth \cite{conf:bwconstraint}, computation limitations \cite{conf:compconstraint}, and generally limited resources. Thus, we envision a new paradigm in the designing and evaluating secure estimation algorithms. Such paradigm requires meeting the real-time requirements and the resource constraints of CPS. Extensive computation and high computational overhead to provide secure estimate estimation are not accepted in our paradigm. To realize this vision, we propose algorithms for secure state estimation in the presence of measurement noise and modeling errors. We show that our algorithms secure the system against physical attacks manifested by additive disruptions on the measurements while keeping in mind the practicality of our proposed algorithms. Up to our knowledge, \sys{} is the first solution that meets the real time requirements of CPS in our considered scenarios.  

%We show are able to detect attacks and estimate the network state in the presence of attacks on the sensor readings.

 An entirely representative application for secure state estimation is the localization and time synchronization problem. With the growing prevalence of wireless devices, it is important to coordinate timing among IoT and to provide contextual information, such as location. Also, position estimation is necessary for different fields such as military \cite{conf:military}, indoor and outdoor localization \cite{conf:indoor}, security surveillance, and wildlife habitat monitoring. Also, maintaining a shared notion of time is critical to the performance of many CPS, Big Science \cite{conf:rabbit}, swarm robotics \cite{conf:formationctrl}, high-frequency trading \cite{conf:ptp}, and global-scale databases \cite{conf:Gspanner}. %Furthermore, l
Localization and time synchronization involves a significant amount of collaboration between individual sensors to perform complex signal processing algorithms. This introduces a considerable amount of vulnerability to the network estimation process. Also, 
the problem is challenging due to its nonlinearity, and due
to the high process and measurement noises. When we deal with time; we are expecting a high effect of modeling errors and measurement noises. 
Therefore, we picked localization and time synchronization as the application driver to illustrate our proposed secure state algorithm. In this work, we revisit the secure estimation problem from a modern viewpoint of sensor networks, to aid in the design of more efficient algorithms with reduced overheads.

%lower bound on the inter-event time are provided. 
% The effectiveness of the proposed approach has been illustrated over a real testbed.
%To realize our vision of a CPS-enabling platform, we propose
 Inspired by the above discussion, we made the following key contributions in this paper:
\begin{itemize}
\item Proposing \sys{} which consists of two secure state estimation algorithms 
%for a network of sensor nodes
suitable for applications in wireless networks with nonlinear time evolution and measurement model. Our solutions are applicable to networks with pairwise measurements.
%\footnote{\be{Currently only algorithms for Sync/Loc are presented, not a general description. Thus, this point is maybe not true.
%\r{I do not see the algorithms limited to sync/loc it is just an application. the big picture can be applied anywhere. It is just adding a state for estimating the attack for every node and use that in EKF and opt. EKF and opt can be used in any domain by just replacing the f,h}}}.
\item Showing that our algorithm is resilient against attacks on the sensor measurements while meeting the real-time requirements of CPS.
\item Applying the proposed strategy in the problem of simultaneous localization and synchronization of spatially distributed nodes in an ad-hoc network. 
\item The proposed algorithms are evaluated on a real testbed using ultra-wideband wireless devices and a quadrotor, representing a network of both static and mobile nodes.
\end{itemize}

 %we aim to apply our event trigger diffusion Kalman filer on D-SLATS \cite{conf:d-slats}.

The rest of the paper is organized as follows: Section \ref{sec:related} gives an overview of the relevant work in the domain. Section \ref{sec:define} defines the secure state estimation problem and the threat model under our study. We define the core concept behind \sys{} in \ref{sec:idea}. We then go through our proposed algorithms in Sections \ref{sec:algoA} and \ref{sec:algoB}. Section \ref{sec:eval} illustrates the experimental setup and evaluates the proposed algorithm on static and mobile network of nodes. Finally, Section \ref{sec:conc} lists some concluding and discussion remarks.

%Therefore, many algorithms are proposed in the literature for 
%\vspace{0.1cm}
\section{Related Work} \label{sec:related}

% We would like to categorize the related work into the following:
In this section, we present the related work on secure state estimation and its application to simultaneous localization and synchronization.

\subsection{Secure State Estimation}
Several recent works have studied the problem of secure state estimation against sensor attacks in dynamical systems. We categorize the work to the following subcategories:

%\vspace{0.1cm}
\subsubsection{Secure Estimation
for Noiseless Systems}
Graph-theoretic conditions for the detectability of attacks for a noiseless system are shown in \cite{conf:noiseless1}. Fawzi et al.  show the impossibility of accurately reconstructing the state of a system if more than half the sensors are attacked \cite{conf:Hamza1}. Pasqualetti et al. \cite{conf:distributed} propose a fully decentralized solution for  attack Identification. 

\subsubsection{Secure Estimation with Bounded non-stochastic Noise}
Satisfiability-Modulo-Theory (SMT) approach is proposed for bounded noise in \cite{conf:bounded1}. Another kind of work was based on brute force search which is not applicable for large-scale systems \cite{conf:bounded2}. A practical solution is proposed in \cite{conf:bounded3-robot} that considers real-time issues and synchronization errors. 
 %which cares about real-time issues such as sampling and actuation jitter, and synchronization errors. 

\subsubsection{Secure state estimation with Gaussian noise}
Observing and recording sensor readings for an amount of time, then repeat them afterward while carrying out an attack is commonly defined as replay attack. Such kind of attack is considered in \cite{conf:gaussian1} where all the existing sensors were attacked. The attacker does not know model knowledge, but he can access and corrupt the sensor data. Another work considered stochastic game anachronism for detecting replay attacks \cite{conf:gaussian2}. Also, denial of service attacks solution under Gaussian noise is proposed in \cite{conf:gaussian3}. Furthermore, false data injection is solved by proposing an ellipsoidal algorithm where attacker’s strategy is formulated as a constrained control problem \cite{conf:gaussian4}.

All the previous work assume \textbf{linear state space model}. Thus, it can not be used with nonlinear state space. Recently, Weber et al. \cite{conf:attackloc} proposed Gordian SMT especially for secure localization which is a non-linear state space. However, they did not propose as a general solution for non-linear secure state estimation of networks with pairwise measurements. Also, during their evaluation, they only considered four links under attack, which is very small number compared to the total number of links. Furthermore, they required about $11$ minutes to be able to provide a secure state estimation which is a huge overhead to recover a network state. Finally, they considered a 2D localization problem only. On the other hand, \sys{} considers the \textbf{non-linear state space model} of networks with pairwise measurements, while taking into consideration the real-time requirements for CPS and assumes powerful attacks comparing to the related work. 

%\vspace{0.1cm}
\subsection{Secure Localization and Time Synchronization}

%Research which was concerned with the 
Researchers attach great importance for secure localization from a long time. For instance, SeRLoc \cite{conf:serloc} proposed a localization scheme based on a two-tier network architecture. Another type of work was based on partial homomorphic encryption to achieve private and resilient localization \cite{conf:proloc}. For more details on secure localization, we refer the reader to \cite{conf:locsurvey1,conf:locsurvey2}. On the other hand, secure time synchronization alone got a lot of attention. TinySeRSync \cite{conf:TinySeRSync}, for example, proposes secure single-hop pairwise time synchronization  using authenticated timestamping. Moreover, authors in \cite{conf:time1} analyze attacks on different time synchronization protocols and propose a secure time synchronization toolbox to counter these attacks. In \cite{conf:cleanslate} a holistic approach to maintain secure time synchronization in a wireless network is proposed, and in \cite{conf:secureSync} pairwise secure synchronization concept is proposed tailored to real-time applications in smart grids. For more details on secure synchronization, we refer the reader to \cite{conf:timesurvey1,conf:timesurvey2}.

Although these two problems, namely secure time synchronization and secure localization, share many aspects in common, they are traditionally treated separately. That results in a lower performance, inefficient use of resources, and non-scalable algorithms. Secnav \cite{conf:secnav} is the only work that covers simultaneous secure localization and time synchronization up to our knowledge at the time of writing. However, Secnav assumes that sensor nodes cannot be compromised by an adversary, and cares only about the attacked communication channel.  In contrast, \sys{} considers attacks which compromise sensor nodes and links.

%\vspace{0.1cm}

\section{Secure State Estimation Problem} \label{sec:define}
In what follows, we discuss our threat model and the utility measures. Then, provide a mathematical formulation of the state estimation problem.

%\vspace{0.1cm}
\subsection{Threat Model} \label{sec:threat}
In the context of attack-resilient state estimation, we have the following entities:
\begin{itemize}
\item \textbf{Estimation Server}: It attempts to securely, efficiently and accurately estimate the state of a network of distributed sensors given that we have measurements and modeling noises. Also, sensors and links are compromised.
\item \textbf{Sensor Nodes}: Entities that make some required sensing in order to help in estimating the network state. The adversary could compromise nodes.
\item \textbf{Adversary}: He modifies the framework with the goal of causing the server to misestimate the state of the network. He can compromise sensor/links to corrupt measurements. Denial of service attack is out of the scope of our paper. Therefore, the receiver will still receive the message from the sender, superimposed by the attacker’s messages. Also, we can look into the problem as compromised messages where man-in-the-middle attack (MITM) is presented to corrupt the delivered messages to the server. 
\end{itemize}

%\vspace{0.1cm}
\subsection{Resilience Metric} \label{sec:metric}
The accuracy of state estimation algorithm is the key for its use in mission-critical systems. Besides, the CPS are often resource-constrained and has to meet some time requirements. To accommodate the above practical constraints while still maintaining the desired resilience level we define the following utility metrics:
\begin{itemize}
\item \textbf{Estimation Error}: The difference between the actual network state (ground truth) and the estimated state.
\item \textbf{Execution Time}: The time needed by the
server to estimate the network state based on the measurements.
\end{itemize}

%More specifically, we consider a number of sensors reporting two types of measurements:
%\begin{itemize}
%item Distance measurements between nodes.
%\item Time stamps of the local clocks of the presented nodes.
%\end{itemize}
%A centralized server is concerned in doing localization and time synchronization for all the nodes in the network. Attack values are added by an adversary to these two types of measurements by compromising either the sensor or the sent message. \\

%\vspace{0.1cm}
\subsection{System and Attack Model}
\label{sec:system}
Consider the following nonlinear time-varying system
\begin{equation}
\begin{split}
 \tilde{x}^{k}_{i+1} &= \tilde{f}( \tilde{x}^{k}_i) + \tilde{n}^k_i\\
y^{k,j}_i &= \tilde{h}^j( \tilde{x}^k_i, \tilde{x}_i^j) + \tilde{v}^k_i + a^{k,j}_i\, ,\\
\end{split}
\label{eq:sysmodel}
\end{equation}
%\footnote{\be{Does it make sense to formulate a second state space model that is used for EKF. The second model includes the attack values to the state: $\tilde{x}_i^j = [x_i^{j\,\text{T}},a_i^{k\,\text{T}}]^\text{T}$ }}
%\footnote{\be{Why do you denote the measurement noise with a single superfix as $v^k_i$, and the attack value by a double superfix as $a^{k,j}_i$.}}
where $x^k_i$ is the state of node $k$ at time $i$. $y^{k,j}_i$ is the measurement sent to node $k$ from the neighborhood node $j \in \mathcal{N}_k$. The process and measurement noise are assumed to be uncorrelated and zero mean white Gaussian noise. They are denoted by $n^k_i$ and $v^k_i$, respectively. $Q_i$ and $R_i$ are the process covariance and the measurement noise matrix at time $i$, respectively. The attack vector $a^{k,j}_i$ is a vector which models how an attacker changes the sensor measurements between node $k$ and node $j$ at time $i$. A non-zero elements in the vector $a$ corresponds to the attacked values on the corresponding sensor; otherwise the measurement is not attacked. Thereby, the additive attack vector $a^{k,j}_i$ can account for both, a malicious node $k$ and a corrupted link $(k,j)$. Moreover, the attack values can be constant or time-varying. %In what follows, we simplify the notation of the attack vector to $a^{k}_i$, as the couple $i$ and $k$ uniquely defines the neighbor $j$ because node $k$ can only communicate with one node at time $i$.

\sys{} considers secure state estimation problem over a network of $N$ nodes indexed by $k\in \{0, \ldots, N-1\}$. The nodes are spatially distributed over some region in space. We say that two nodes are connected if they can communicate directly with each other. The neighborhood of a given node $k$ is denoted by the set $\mathcal{N}_k$.
 %Our node represents some sensors and estimator.
 %conf:d-slats
\section{\sys{} Core Idea} \label{sec:idea}

%\section{\sys{} Core Idea} \label{sec:idea}
Most of the related work on secure state estimation mitigates the influence of attacks as described in Section \ref{sec:related} by identifying the attacked links and excluding them from the estimation process. This is a highly complex combinatorial problem, and even the convex relaxation techniques are not suitable for real-time applications. So, let's go through \sys{} core idea that helps us to solve the problem while considering the practical requirements.

To understand the idea behind \sys{}, let's consider the following toy example which is representative of the core idea behind our solution. Consider a network of $N$ nodes, where each node has a link with its neighbor. Now, let's consider Figure \ref{fig:toy}, node$_1$ will communicate with node$_2$ at $t_1$. Then node$_1$ will communicate with node$_2$ and node$_3$ at time $t_2$ and $t_3$, respectively. Now, let's go back to the general model of the secure state estimation which is shown in Equation \ref{eq:sysmodel}. According to that model, we have three attacks which are $a^{2,1}_1$, $a^{3,1}_2$, and $a^{4,1}_3$ on the three links. 
Related work tried to solve at each time instant $t$ the secure state estimation problem while considering the number of possible attacks (unknowns) equals the number of links. It ends up in a very complicated problem with the number of unknowns equals the number of connections.
%To illustrate the principle followed in this work, we want to emphasize the particular structure of the system model, i.e., of the pairwise measurements taken per time step $i$. Therefore, consider a network application as depicted in Fig.~\ref{fig:toy}, Where different network nodes communicate with each other. This is done in a way that per time interval $i$ only measurements over a single link are taken at time $t_i$. Note that due to this principle it is possible to denote the attack $a_i^{k,j}$ only by $a_i^{k}$, i.e., neighbor $j$ is uniquely defined by time step $i$ and node $k$.

%% here------------
In contrast, we look into the problem from a different perspective inspired by the real-life application's constraints. At each time instant, $t$ each node can only communicate with one neighbor and this is the practical scenario in any CPS. For instance, measuring the distance using time of the flight of two or three messages between two nodes is a type of communication \cite{conf:d-slats}. Thus, at time $t_1$ in Figure \ref{fig:toy}, why should we care about estimating $a^{3,1}_1$?. What about renaming the three attacks on the figure to $a^{1}_1$, $a^{1}_2$, and $a^{1}_3$ as the couple time instant $t_i$ and the node index $k$ uniquely defines the neighbor $j$ because node $k$ can only communicate with one node at time $t_i$. Thus, \sys{} looks into the attack that is related to node$_1$ as $a^{1}_t$ and deals with it as time-varying value. With this paradigm shift, we are able to have the number of attack at each time instant equals to the number of nodes, instead of the number of links. Therefore, \sys{} drastically reduces the computational complexity. We will show the performance of \sys{} while attacking \textbf{all} the links with time-varying attacks. State estimation while attacking all the links is a hard problem and could not be solved in any of the related work up to our knowledge. 

We now utilize this idea to change the general model in \eqref{eq:sysmodel}. We include the attack value of this time step to the state of the node initiating the measurement. Following this procedure, the state of node $k$ would be extended to ${x}_i^k = [\tilde{x}_i^{k\,\text{T}}, a_i^{k\,\text{T}}]^\text{T}$. This yields a modified system model with ${x}_{i+1}^k = {f}({x}_i^k) + {n}_i^k$ and $y_{i}^k ={h}^j({x}_i^k,x_i^j) + {v}_i^k$. In this way, we decrease the number of unknowns at each time step. Instead of being proportional to the number of links, they are now only proportional to the number of nodes. We should emphasize that this is just a modeling trick and all the links can be under attack with different values.

%In contrast, in this work, we propose a paradigm shift to reduce the computational complexity drastically. As the measurement model in \eqref{eq:sysmodel} accounts only for measurements involving a subset of nodes per time step $i$, we include the attack value of this time step to the state of the node initiating the measurement. Following this procedure, the state of node $k$ will be extended to ${x}_i^k = [\tilde{x}_i^{k\,\text{T}}, a_i^{k\,\text{T}}]^\text{T}$. 
%This yields a modified system model with ${x}_{i+1}^k = {f}({x}_i^k) + {n}_i^k$ and $y_{i}^k ={h}^j({x}_i^k,x_i^j) + {v}_i^k$. In this way, we decrease the number of unknowns at each time step. Instead of being proportional to the number of links, they are now only proportional to the number of nodes. Moreover, we do not need to solve a combinatorial problem and thus can significantly reduce the computational complexity.

%One challenging point in our approach is how to set the process update function when it comes to attacker behavior. %We choose to have $a_i^k = a_{i+1}^k$. 
A valid question would be how to model the time-evolution in the state-space model for unpredictable static or time-varying attacks. We choose to have $a_i^k = a_{i+1}^k$, and the variance of the attack entries of the modeling noise ${n}_i^k$ would account for the time-varying aspect of the attack. We would like to define two important concepts associated with measurement values: 

\begin{itemize}
\item Outliers: These measurement values lie outside the main expected range. For example, it would be the measured distance between two nodes that is much larger than diagonal of the localization area. Also, if the measured speed of moving node is higher than the hardware capabilities of the nodes, then this speed value would be considered an outlier. These Outliers can be easily detected and rejected by setting some thresholds on the accepted values.

\item Attacks: the Smart attacker would corrupt the sensor reading with new values that make sense. Thus, attacks are values that can't be removed by outliers' detectors. Attacked speed value, for example, would be within the capabilities of the node motors.
\end{itemize}

\sys{} makes use of this concept to simplify the complexity of secure state estimation algorithms. We care about designing secure state estimation algorithm which would be combined with outliers rejectors. Thus, attacks can only be in a limited interval which can be represented by the modeling variance. We used this concept in our previous work \cite{conf:d-slats,JacobiCDC17} where we use stationary model for process update function for tracking a flying quadrotor. As long as the quadrotor moves in the range of the modeling noise, D-SLATS would be able perfectly to localize it. At the end of the day; the speed of the quadrotor is limited by the hardware capabilities which are public information. So why we do not make use of this information to simplify the estimation algorithm?

%The D-SLATS work and to related work.\cite{conf:d-slats} Inspires our idea, where they use stationary model for process update function, and it works well for a flying quadrotor. The powerful capability of accounting for the modeling noise is core concept behind that. As long as the quadrotor moves in the range of the modeling noise, D-SLATS would be able perfectly to localize it. Therefore, our model would be able to account for time-varying attack and this will be shown in practical experiments. Also, from a practical point of view, the attack should be within the acceptable sensor measurements range; otherwise, it can be easily detected. Thus, determining the noise covariance entries related to attack ${n}_i^k$ would depend on the application and sensor ranges.

%2- 

In what follows, this approach will be applied to simultaneous localization and synchronization. It will be seen that the system performance is greatly improved by this approach while the computational complexity is significantly reduced compared to the related work.
%\b{ Detecting and mitigating attacks on sensor data is, in general, a combinatorial problem \cite{conf:comb} which has been previously addressed by brute force search or convex relaxations \cite{conf:Hamza1,conf:comb}. These algorithms can not meet the real-time requirements of the CPS due to its complexity. Thus, we consider a simplified formulation. Instead of considering the number of attacks equals the number of links, we consider the number of attacks equals the number of nodes at each time instance $i$, as each node would perform one communication at each time instance i. Such simplification leads to significant improvement in the system performance and does not limit our algorithms to constant attacks. }

%\vspace{0.1cm}
\section{\sys{} in Real Life Application} \label{sec:app}
Without loss of generality, we choose to apply \sys{} on simultaneous localization and time synchronization estimation problem while some nodes/links are under attack. It a quite representative example of secure state estimation problem where we have modeling noise of the clocks and the dynamics of moving nodes in conjunction with measurements noise. Our sensors are able to get the time difference and the distance between two each other. More specifically, we consider three types of measurements which are distinguished by the number of messages exchanged between a pair of nodes. The measurement vector at node $k$, from node $j \in \mathcal{N}_k$ has the following form%\joao{This is not quite right because you now have the ${ao}$ ${ad}$ both as noise in (3) and as part of the state in (5).}
\begin{eqnarray}
y^{k,j}_i &= &\begin{bmatrix}
d^{k,j}_i +{ao}^k_i\\
r^{k,j}_i + {ad}^k_i\\
R^{k,j}_i +{ad}^k_i
\end{bmatrix}\, ,
\end{eqnarray}
where $d^{k,j}_i$ represents the \textbf{counter difference} at time $i$ denoted by $d^{k,j}_i$ which is the measurement of the difference between the clocks of each node. $r^{k,j}_i$, on the other hand, represents a noisy measurement due to frequency bias discrepancies between $k$ and $j$ which is formally represented by \textbf{single-sided two-way range}. Finally, $R^{k,j}_i$ is another \textbf{distance measurement} between nodes $k$ and $j$ based on a trio of messages between the nodes at time $i$. This is a more accurate estimate than $r^{k,j}_i$ due to mitigation of frequency bias errors from the additional message. It is formally called \textbf{double-sided two-way range}. For more details about the three types of measurements, we encourage the reader to check \cite{conf:d-slats}. However, these measurements are corrupted by some attacks on counter difference and distance, namely, ${ao}^k_i$ and ${ad}^k_i$, respectively. We would like to note that subset of these measurements may be used rather than the full set, i.e., we can have experiments involving just $r^{k,j}_i$, $R^{k,j}_i$, $d^{k,j}_i$ or any combinations. Also, we do not put any limitation on the way of calculating these measurements. So, no special requirements on the used hardware. 
%Time stamps $t_0(i)$ through $t_5(i)$ denote the locally measured transmission (TX) and reception (RX) times stamps, and $T_{RSP}(i)$ and $T_{RND}(i)$ define, respectively, the response and the round-trip durations between the appropriate pair of these timestamps. The propagation velocity of radio is taken to be the speed of light in a vacuum, denoted by $c$. 

Thus, we are concerned with a state vector which consists of three dimensional position vector $\bm{p}^k_i$, clock time offset $o^k_i$, and clock frequency bias $b^k_i$ for every node. We adopt a convention where both $o^k_i$ and $b^k_i$ are described with respect to the global time clock which is usually the clock of master node, which can be any node. Also, there are attacks on the distance measurements, which is denoted by ${ad}^k_i$, and the attack on offset measurement values ${ao}^k_i$ which are added by an attacked node. In summary our state vector is as following:%\joao{If ${ao}^k_i$ and ${ad}^k_i$ are part of the state, they cannot be seen as measurement noise. What will happen is that the Kalman filter will try to estimate these quantities and generally succeed if the attacker keeps them constant. However, this will not work if the attacker keeps changing them.}
% $x^k_i =\left[{\bm{p}^k_i}^T,\: o^k_i,\: b^k_i\right]^{T}$, 
\vspace{-0.2cm}
\begin{eqnarray}
x^k_i &= &\begin{bmatrix}
{\bm{p}^k_i}^\text{T}& o^k_i& b^k_i& {ao}^k_i& {ad}^k_i
\end{bmatrix}^{\text{T}}
\end{eqnarray}
%\vspace{-0.2cm}
We define the full network state $x_i$ at master time as the concatenation of all states from the N participating devices at time step $i$:\\[-0.6cm]
%\vspace{-0.2cm}
\begin{eqnarray}
x_i &= &\begin{bmatrix}
x^{0\,\text{T}}_i& x^{1\,\text{T}}_i& ..&x^{N-1\,\text{T}}_i
\end{bmatrix}^{\text{T}}
\end{eqnarray}
\vspace{-5.8mm}

\vspace{0.1cm}
\section{\sys{} Attack Resilient State Estimators}

\subsection{\algone{} Algorithm} \label{sec:algoA}

The core idea behind \algone{} is to make use of Extended Kalman filter (EKF) in order to estimate the network state beside the introduced attacks. We periodically apply corrections when new measurements are
received. We assume the clock parameters evolve according to the first-order affine approximation of the following dynamics $o^k_{i+1} = o^k_{i} + b^k_{i} \delta_t$ and $b^k_{i+1} = b^k_i$, where $\delta_t:=t_M(i+1)-t_M(i)$ given that $t_M$ is the root node time which is the global time. The process update is responsible for evolving the state estimates for all N devices—clock offsets, frequency biases, and device position. Therefore, we can write the process update function as following:
\begin{eqnarray}
f( x^k_i) &= &\begin{bmatrix}
\bm{p}^k_i\\
o^k_i+b^k_i\delta_{t}\\
b^k_i\\
{ao}^k_i\\
{ad}^k_i
\end{bmatrix}
\end{eqnarray}

Note that this simple process model can still be used to capture node state dynamics and changing attack values simply by setting the covariance of the corresponding state value adequately high to account for high change rate. We will show that our module can work with high rate changing attack and moving nodes as well. On the other hand, The measurement update step relates the state variables $x^{k}_{i}$ to a measurement vector $y^{k,j}_i$ between two devices. Each of our measurements depends on the  of local clock values, distances between two devices and the amount of the introduced attack on time and distance. For any given node $k$ in the network, the node-specific measurement model with respect to a
second node $j$ is given as:
\begin{align}
h^j(x^k_i,x^j_i)&=\left[\begin{array}{c}
%\left(o^j_i-o^k_i\right) - {ao}^k_i +  \left\Vert \bm{p}^{j}_i-\bm{p}^{k}_i\right\Vert _{2}/c\\
\left(o^j_i-o^k_i\right) + {ao}^k_i\\
(1+b^k_i)\left\Vert \bm{p}^{j}_i-\bm{p}^{k}_i\right\Vert _{2} + {ad}^k_i \\
(1+b^k_i)\left\Vert \bm{p}^{j}_i-\bm{p}^k_{i}\right\Vert _{2}+ {ad}^k_i \\ %nce between 2nd and 3rd line?
\end{array}\right]
\label{eq:measmodel}
\end{align}

%\begin{align}
%h^j(x^k_i)&=\left[\begin{array}{c}
%\left(o^j_i-o^k_i\right) - {ao}^k_i +  \left\Vert \bm{p}^{j}_i-\bm{p}^{k}_i\right\Vert _{2}/c\\
%\left(o^j_i-o^k_i\right) + {ao}^k_i\\
%\left\Vert \bm{p}^{j}_i-\bm{p}^{k}_i\right\Vert _{2} + {ad}^k_i +\frac{c}{2}\left(b^{j}_i-b^{k}_i\right)T_{RSP1}\\
%\left\Vert \bm{p}^{j}_i-\bm{p}^k_{i}\right\Vert _{2}+ {ad}^k_i + c\cdot \tilde{R}^{k,j}_i\\
%\end{array}\right]
%\label{eq:measmodel}
%\end{align}

%\joao{Up to here, this section mostly describes the model, so it should probably be moved to the previous section.}

The time synchronization and localization problem aims to estimate the clock parameters and the 3D position using the available measurements and the model of the system. We denote by $\hat{x}_{i}$ the estimate of $x_i$ where we seek to minimize the mean-square error $\text{E} \| x^i-\hat{x}_i\|^2$. Given this, we design an EKF to get a secure state estimate over time, periodically applying corrections when new timing and range measurements are received. Our EKF only updates upon arrival of a new measurement (message) exchanged between two or more nodes. The EKF steps are:
%We can summarize the steps in the following items:

\begin{itemize}
\item{\textbf{Receive a new measurement:}} The server obtains a new measurement at filter step $i$ which is generated by a message exchange between devices $k$ and $j$. By convention, communication is always defined to terminate at device $j$. One of the nodes maybe under attack and corrupt the sent measurement.
\item{\textbf{Calculate $\delta_{t}$:}} The server maintains an estimate of master time $t_M(i)$ as well as time offset and frequency bias estimates for every node. This gives the capability to calculate the time elapsed between measurements in the master time reference. %$\delta_t:=t_M(i+1)-t_M(i)$
\item{\textbf{Time update:}} Update the current secure estimates for the location and time of each node while taking the attack effect out of the measurement.
%\item{\textbf{Measurement update:}} Convert the measurement to an observation of ranges, frequency biases, clock offsets, attack estimates and correct the secure state.
\end{itemize}

\subsection{\algtwo{} Algorithm} \label{sec:algoB}
The second algorithm for secure state estimation is based on the Maximum Likelihood Estimator (MLE). By using the model of the system and the knowledge of how the unknown attack signal affects the measurements, we formulate an optimization problem as
\begin{equation*}
\text{arg} \hspace{-0.2cm} \min_{ x_i^k, a_i^{k}\, \forall k,j} 
\sum_{\begin{array}{c} \rule{1mm}{0mm}\\[-5.2mm]{\scriptstyle j,k}\\[-1.5mm]{\scriptstyle i' \in \{i-L,i\}} \end{array}} %_{j,k} 
\hspace{-0.4cm}
\Big(
\|h^j( x_{i'}^k,x_{i'}^j)-y_{i'}^{k,j}\|^2
-\lambda \| a_{i'}^{k}\|_1
\Big)\, ,
\end{equation*}
where a time window comprising $L$ measurements is used for minimization. The L1-norm on the attack values is added to quadratic minimization criteria for more robust attack estimation. The resulting method is summarized in Algorithm \ref{alg:SecOpt}. The algorithm is centralized in the sense that all the exchanged messages are transmitted to a centralized server where the computation and optimization are solved. While the minimization is initialized for $i=0$ with the state $x_\text{init}=0$ for numerical optimization, run repeat the it multiple times by setting $x_\text{init}$ to the previously estimated state.
%Note that the added L1 norm of the attacks limits
%that estimates the value of the state vector that best describes the observed measurements.

%To the classical MLE we add a regularization term, inspired by the LASSO regression analysis. It is known that redundancy in the measurements is necessary for the uniqueness localization of systems under attack \cite{conf:attackloc}. With the assumption that most of the measurements are not gonna be corrupted by an attacker, we add a L1 norm regularization component that promotes sparsity when estimating the attack vector.

\begin{algorithm}
\caption{\algtwo{} \label{alg:SecOpt}}
\begin{flushleft}
%\textbf{Algorithm \algtwo{}}\\
Set $i = 0$, and $\hat{x}_{0}= x_\text{init}$. At the centralized server:\\
\textbf{Step 1}: Collect enough measurements between nodes  according to the window size.\\
\textbf{Step 2}: Solve the centralized optimization problem with solver initial condition set to $\hat x_{i}$:\\
\begin{align}
\left[\hat{o}^j_{i+1}~\hat{\bm{p}}^j_{i+1}~\hat{ao}^j_{i+1} \right] =
&\argminB_{o^j_i,\bm{p}^j_i,ao^j_i} \sum_{j \in \mathcal{N}_k} (d^{k,j}_{i}-( o^k_{i} - {o}^j_{i} ) \nonumber\\
&- \left\Vert \bm{p}^k_{i}-{\bm{p}}^j_{i}\right\Vert _{2}/c  - {ao}^j_i )^2 +\lambda \| {ao}_i^{j}\|_1
%\label{eq:opt1} 
\nonumber\\
\left[~\hat{\bm{p}}^j_{i+1}~\hat{ad}^j_{i+1} \right] = &\argminB_{b^j_i,\bm{p}^j_i,ad^j_i} \sum_{j \in \mathcal{N}_k} ( R^{k,j}_{i} - \left\Vert \bm{p}^k_{i}-{\bm{p}}^j_{i}\right\Vert _{2} \nonumber\\
&-  {ad}^j_i)^2 +\lambda \| {ad}_i^{j}\|_1 \nonumber
%\label{eq:opt3}
\end{align}
\textbf{Step 3}: Increment $i$ by 1 and go back to Step 1

%\textbf{Step 3}: Combine the results from the optimization step over all nodes.
\end{flushleft}
\end{algorithm}

%In order to improve the estimate and mitigate the measurement errors and have redundancy, we collect the measurement during a fixed window time and we run the algorithm multiples times feeding in the previous estimate as the initial condition for the optimization solver.

%\begin{align}
%\left[\hat{o}^j_{i+1}~\hat{\bm{p}}^j_{i+1}~%\hat{ao}^j_{i+1} \right] =
%&\argminB_{o^j_i,\bm{p}^j_i,ao^j_i} \sum_{j \in %\mathcal{N}_k} (d^{k,j}_{i}-( o^k_{i} - \hat{o}^j_{i} ) %\nonumber\\
%&- \left\Vert \bm{p}^k_{i}-\hat{\bm{p}}^j_{i}\right\Vert _{2}/c  - \hat{ao}^j_i )^2
%\label{eq:opt1} 
%\nonumber\\
%\left[\hat{b}^j_{i+1}~\hat{\bm{p}}^j_{i+1}~\hat{ad}^j_{i+1} \right] = &\argminB_{b^j_i,%\bm{p}^j_i,ad^j_i} \sum_{j \in \mathcal{N}_k} ( R^{k,j}_{i} - \left\Vert \bm{p}^k_{i}-\hat{\bm{p}}^j_{i}%\right\Vert _{2} \nonumber\\
%&-  \hat{ad}^j_i)^2 \nonumber\\
%c\cdot \tilde{R}^{k,j}_i -
%where \left\Vert  \hat{ao} , \hat{ad}  \right%\Vert_{\ell1} < K  \nonumber
%\label{eq:opt3}
%\end{align}

%\input{Sections/6-algoB_theory.tex}
\section{Evaluation} \label{sec:eval}

%\begin{figure}[t]
%\centering
%\includegraphics[width = 0.5\textwidth]{Figures/testbed_v2.jpg}
%\caption{(a) CrazyFlie 2.0 helicopter equipped with the very same Decawave ultra-wideband radio, (b) Ceiling-mounted anchor node
%, and (c) Custom ranging anchor circuit board}
%\label{fig:testbed_v2}
%  \vspace{-0.4cm}
%\end{figure}

% 
% \begin{figure*}
% \centering
% \begin{minipage}[t]{0.30\textwidth}%
% \includegraphics[height=1.75in]{Figures/ntb_anon.jpg}
% \caption{Custom anchor node with ARM Cortex M4 processor and UWB expansion.}
% \label{fig:ntbanon}
% \end{minipage}\hspace{5mm}
% \begin{minipage}[t]{0.30\textwidth}%
% \centering
% \includegraphics[height=1.75in]{Figures/ntb_ceil.jpg}
% \caption{Ceiling-mounted anchor with DW1000 UWB radio in 3D-printed enclosure.}
% \label{fig:ntbceil}
% \end{minipage}\hspace{5mm}
% \begin{minipage}[t]{0.30\textwidth}%
% \centering
% \includegraphics[height=1.75in]{Figures/quad.jpg}
% \caption{CrazyFlie 2.0 quadrotor helicopter with DW1000 UWB expansion.}
% \label{fig:quad}
% \end{minipage}
% \end{figure*}

\subsection{Experimental Setup}

We evaluated the performance of the proposed secure state estimation algorithms in a custom ultra-wideband RF testbed based. The overall setup is described in details in the Appendix and shown in Figure \ref{fig:full_sys}. 
\subsection{Experiments}
We demonstrate \sys{} state estimation algorithms in simultaneous localization and time synchronization of wireless sensor nodes under attack while keeping on minds the resilience metrics as defined in Subsection \ref{sec:metric}. This attack can be on the communication link or/and the whole sensor node. We are going to evaluate our proposed algorithms under different attack cases for static and mobile nodes.

\subsection{Case Study: Static Nodes} \label{sec:static}

Nodes are placed in 8 distinct locations around the 10 $\times$ 9 $m^2$ area, roughly in the positions indicated by Figure \ref{fig:full_sys}. The goal of our algorithm is to accurately and securely estimate the positions of all network devices relative to each other, as well as, the relative clock offsets. This relative localization (graph realization), is a well-researched field. Local minima, high computational complexity, and restrictions on graph rigidity are the main challenges in graph realization problems \cite{grp_aspnes,grp_jackson}. \sys{} does not put restrictions on the connectivity of the graph as we will show. An original extended Kalman filter (\origekf{}) is implemented as a baseline for comparison to show the criticality of estimating the attacks values in CPS.

\subsubsection{Secure Position Estimation} 
We would like to start with defining the attacker strategy to corrupt the network state estimation. The common way in secure estimation field is to consider a uniform random number or constant values as attacks value without putting constraints on the rate of change of the attacks values. Neglecting the convergence time of the algorithm has catastrophic consensuses. Unlike the previous work, we consider three types of attacks generator statistics. Also, we consider the attacker switch between different parameters during an attack session. Function \textbf{1} in the Appendix shows a Pseudocode for our attacker function. We consider Uniform, Normal and Pareto Distributions as type 1, 2, and 3 attacks, respectively. Note that generating the attack values according to Function \textbf{1} simulates time-varying attack values that are different for all links. Also, we should highlight the following points:

%\ref{sec:appendix} 
\begin{itemize}
\item Every new measurement gets different attack value, i.e, the Function \textbf{getDistanceAttackValue} is called every time instance for every new measurements.
\item All the measurements under attack. 
\item The attacker changes the parameters of each type from period to period as shown in the Pseudocode. Such an attack is stronger than using fixed statistics random generator.
\item The attacks values must results in a location within the localization area, as a smart attacker strategy. Otherwise, the attack can be easily detected with a simple threshold mechanism.
\end{itemize}

%\belowcaptionskip=-10pt

\begin{figure*}[tbp]
%\vspace{-0.05cm}
    \centering
    \begin{tabular}{ p{0.29\textwidth}  p{0.29\textwidth}  p{0.29\textwidth}}
        \resizebox{0.29\textwidth}{!}{
            \begin{subfigure}[h]{0.29\textwidth}
      \centering
        \includegraphics[scale=0.45]{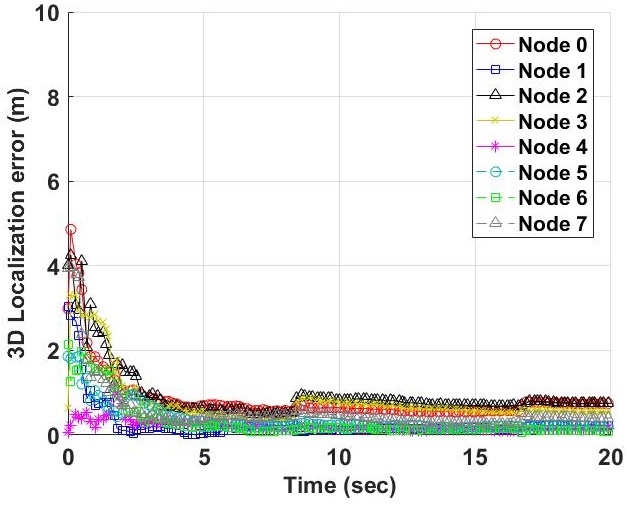}
        \caption{\algone{}}
        \label{fig:type1_secekf}
    \end{subfigure}
       } 
   &
   \resizebox{0.29\textwidth}{!}{
            \begin{subfigure}[h]{0.29\textwidth}
      \centering
        \includegraphics[scale=0.45]{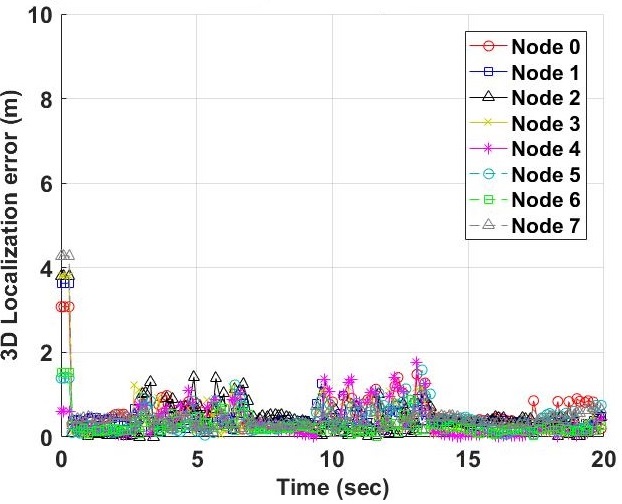}
        \caption{\algtwo{}}
        \label{fig:type1_secopt}
    \end{subfigure}
      }
    &
   \resizebox{0.29\textwidth}{!}{
         \begin{subfigure}[h]{0.29\textwidth}
      \centering
        \includegraphics[scale=0.45]{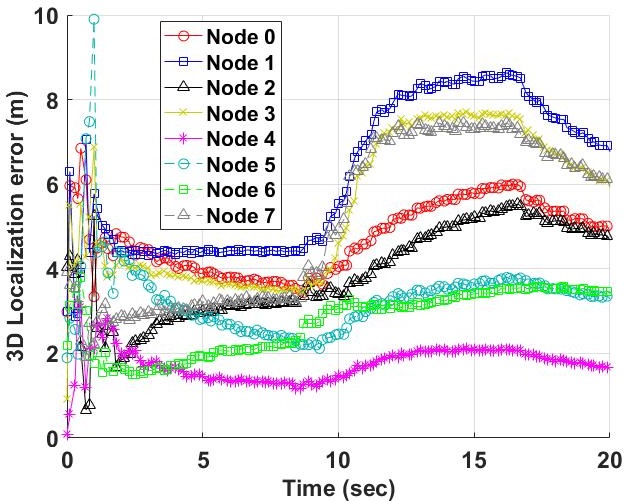}
        \caption{\origekf{}}
        \label{fig:type1_orgekf}
    \end{subfigure}
    }
    \\
    \resizebox{0.29\textwidth}{!}{
            \begin{subfigure}[h]{0.29\textwidth}
      \centering
        \includegraphics[scale=0.45]{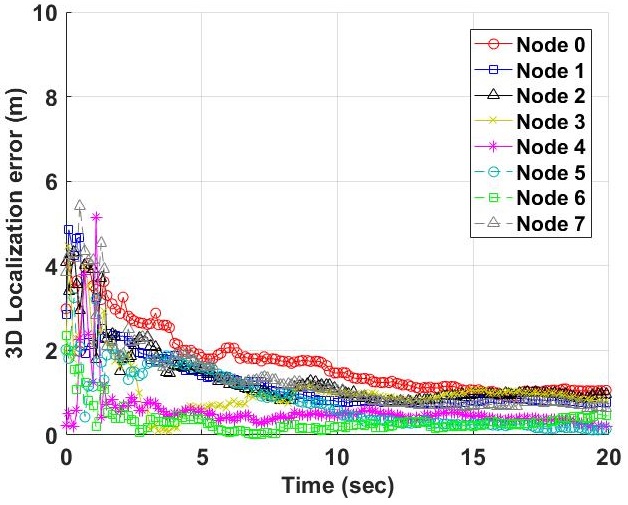}
        \caption{\algone{}}
        \label{fig:type2_secekf}
    \end{subfigure}
       } 
   &
   \resizebox{0.29\textwidth}{!}{
            \begin{subfigure}[h]{0.29\textwidth}
      \centering
        \includegraphics[scale=0.45]{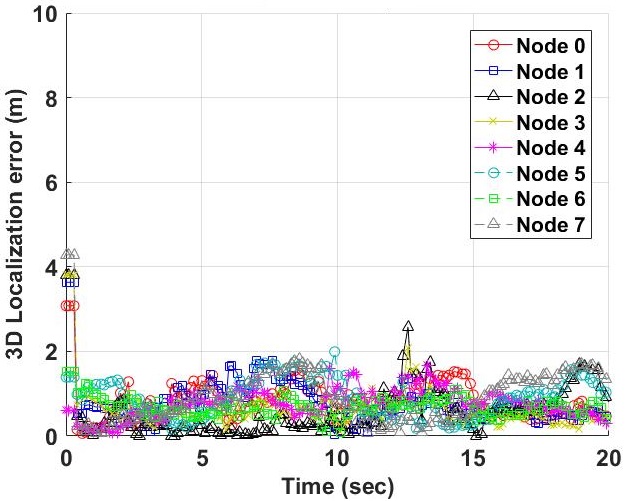}
        \caption{\algtwo{}}
        \label{fig:type2_secopt}
    \end{subfigure}
      }
    &
   \resizebox{0.29\textwidth}{!}{
            \begin{subfigure}[h]{0.29\textwidth}
      \centering
        \includegraphics[scale=0.45]{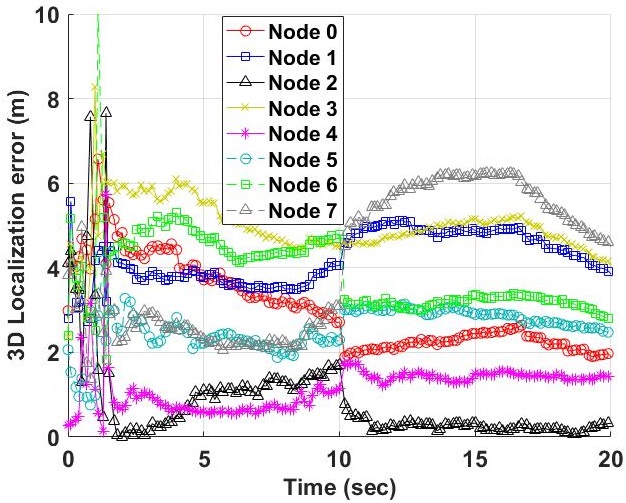}
        \caption{\origekf{}}
        \label{fig:type2_orgekf}
    \end{subfigure}
    }
   \\ 
    \resizebox{0.29\textwidth}{!}{
            \begin{subfigure}[h]{0.29\textwidth}
      \centering
        \includegraphics[scale=0.45]{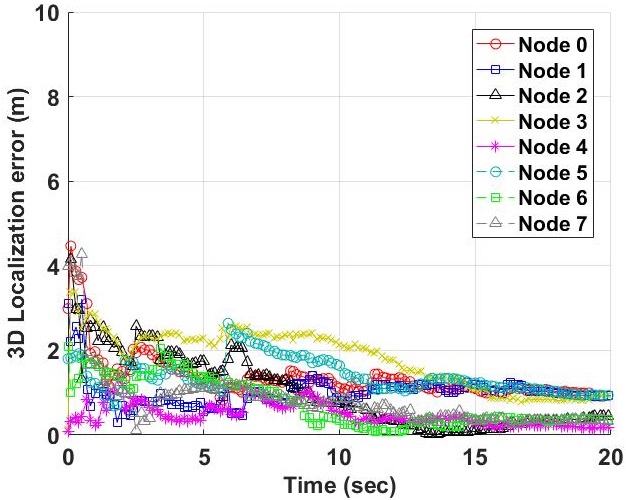}
        \caption{\algone{}}
        \label{fig:type3_secekf}
    \end{subfigure}
       } 
   &
   \resizebox{0.29\textwidth}{!}{
            \begin{subfigure}[h]{0.29\textwidth}
      \centering
        \includegraphics[scale=0.45]{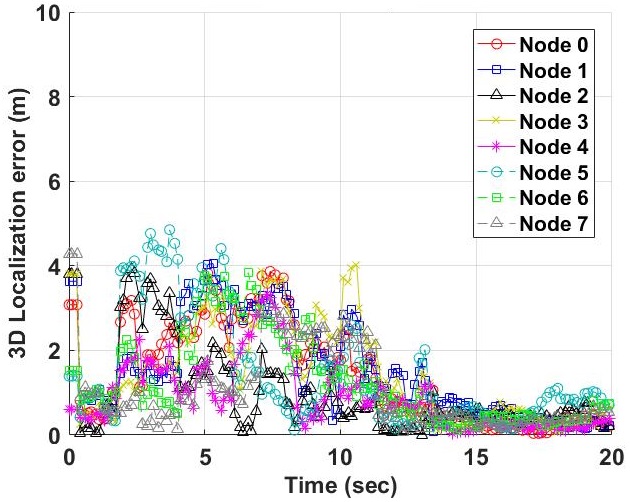}
        \caption{\algtwo{}}
        \label{fig:type3_secopt}
    \end{subfigure}
      }
    &
     \resizebox{0.29\textwidth}{!}{
            \begin{subfigure}[h]{0.29\textwidth}
      \centering
        \includegraphics[scale=0.45]{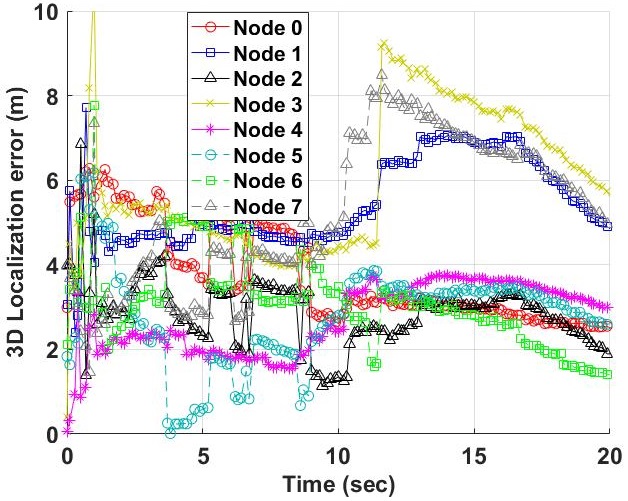}
        \caption{\origekf{}}
        \label{fig:type3_orgekf}
    \end{subfigure}
    }
   \\ 
  \end{tabular}
\caption{Localization error for the eight nodes where all the measurements are under attack with different values. Sub-figures (a), (b) and (c) correspond to Type 1 attack which is time varying uniform random number. Sub-figures (d), (e) and (f) correspond to Type 2 attack is time varying normal distribution. Sub-figures (g), (h) and (i) correspond to Type 3 attack which is following time varying Pareto Distribution. Function \textbf{getDistanceAttackValue} shows the attacker behavior for corrupting the measurements $R_ij$ and $r_ij$.}% The three algorithms are running in parallel on the same set of attacks values to have a fair comparison.}
    \label{fig:disatt}%\vspace{-4mm}
      \vspace{-0.2cm}
\end{figure*}

%\newcommand{\algone}{SecEKF} 
%\newcommand{\algtwo}{SecOPT} 
%\newcommand{\origekf}{OrigEKF}
The proposed secure algorithms find relative positions. Therefore, we first superimpose the estimated positions onto the true positions of each node by use of a Procrustes transformation \cite{procrustes}. Specifically, the estimated network topology as a whole is rotated and translated without scaling, until it most closely matches the true node positions. Once transformed, the error of a given node's position is defined as the $\ell_2$ norm of the transformed position minus the true position. Figure \ref{fig:disatt} shows a twenty seconds snapshot of the performance after running the three types of attacks. We can notice the jump in the \origekf{} behaviors due changing the attack statistics. However, \sys{} is resilient against such type of attacks. Table \ref{tab:loc_err} summarizes the localization error for every node using the three algorithm. \algone{} outperforms \algtwo{} in the three types of attacks.

%\begin{figure}[tb]
%\centering
%\includegraphics[width = 0.45\textwidth]{Figures/typ1-20sec.jpg}
%\caption{Type 1.}
%\label{fig:typ1-20sec}
%\end{figure}

\begin{table*}[tbp]
\centering
\normalsize
\begin{tabular}{c|l||c|c|c|c|c|c|c|c||c|c}
% &\multicolumn{3}{c}{Single-sided ranging}&\multicolumn{3}{c}{Double-sided ranging}\\
 
& Algorithm & node 0 &node 1 & node 2& node 3&node 4&node 5&node 6& node 7& mean & std \\
\hline\hline
\multirow{3}{*}{\begin{turn}{90}Type 1\end{turn}}
&\algone{}     &  0.655 & 0.177 & 0.720 & 0.572 & 0.154 & 0.229 & 0.134 & 0.387 &  0.378 & 0.240\\
&\algtwo{}    & 0.607 & 0.668 & 0.650 & 0.606 & 0.424 & 0.553 & 0.400 & 0.713  &0.578 & 0.113 \\
&\origekf{}    & 4.744 & 6.588 & 4.251 & 5.787 & 1.665 & 3.106 & 3.048 & 5.711   & 4.362 & 1.671 \\
\hline
\multirow{3}{*}{\begin{turn}{90}Type 2\end{turn}}
&\algone{}     &  1.343 & 0.921 & 1.016 & 0.864 & 0.390 & 0.675 & 0.323 & 0.942 &   0.809 & 0.336\\
&\algtwo{}  & 0.874 & 0.973 & 0.767 & 0.832 & 0.767 & 0.984 & 0.675 & 1.288 & 0.895 & 0.190\\
&\origekf{} &  2.546 & 4.134 & 0.718 & 4.579 & 1.213 & 2.531 & 3.479 & 4.465 & 2.958 & 1.462\\
\hline
\multirow{3}{*}{\begin{turn}{90}Type 3\end{turn}}
&\algone{}      &  1.107 & 0.960 & 0.849 & 1.596 & 0.444 & 1.314 & 0.633 & 0.627 &    0.941 & 0.386 \\
&\algtwo{}  & 1.631 & 1.692 & 1.210 & 1.635 & 1.053 & 1.769 & 1.476 & 1.267 &  1.467 & 0.260 \\
&\origekf{}  &  3.206 & 5.332 & 2.450 & 5.901 & 2.856 & 2.649 & 2.825 & 5.409 &  3.829 & 1.448\\ 
\end{tabular}
\caption{Localization error for the eight nodes where all the measurements are under Type 1, 2, and 3 attack with different values. Function \textbf{getDistanceAttackValue} shows the attacker behavior for corrupting the measurements. The three algorithms are running in parallel on the same set of attacks values to have a fair comparison.}
\label{tab:loc_err}
  \vspace{-0.5cm}
\end{table*}

\subsubsection{Secure Time Synchronization}

We considered two attacker strategies in order to corrupt the network state estimation in term of time, for the space constraint. We define Type 4 attack where each node adds a constant value in order $\sim100 \mu$ seconds to measurement $d_ij$. While, Type 5 attack considers adding uniform random values in order $\sim100 \mu$ seconds to measurement $d_ij$. The reason behind choosing these types of attacks is discussed in the Appendix. We should highlight that different attacked nodes add different attack values for both Type 4 and Type 5. As mentioned before, we choose node 0 to be the reference node, i.e, $b^0_{i}:=0$ and $o^0_{i}:=0~~\forall i$. Table \ref{tab:sync_err} shows the synchronization errors for all nodes with respect to node 0. We also assume that the reference node is free of attack which is not the same assumption in the secure localization scenario. In order to test the secure time synchronization with minimum uncertainty in our testing mechanism, we follow the testing strategy in \cite{conf:uncer}, and \cite{conf:pulsesync}.

\algone{} comes with the best performance, then \algtwo{} for both attack types. We should note that the synchronization errors in Table~\ref{tab:sync_err} are reported for a fully connected network. \algone{} algorithm achieved $0.03$ and $0.25$ $\mu$ seconds for Type 4 and Type 5 attack, respectively. On the other hand, \algtwo{} algorithm achieved $0.34$ and $9.10$ $\mu$ seconds for Type 4 and Type 5 attacks, respectively. Neglecting the secure estimation techniques leads to catastrophic consequences as shown for \origekf{} performance against Type 4 and Type 5 attacks. The synchronization error is in order of attacks values, namely, $121$ and $132$ $\mu$ seconds for Type 4 and Type 5 attack, respectively.

\begin{table*}[tbp]
\centering
\normalsize
\begin{tabular}{c|l||c|c|c|c|c|c|c||c|c}
% &\multicolumn{3}{c}{Single-sided ranging}&\multicolumn{3}{c}{Double-sided ranging}\\
 
&Algorithm & node 1 & node 2& node 3&node 4&node 5&node 6& node 7& mean & std \\
\hline\hline
\multirow{3}{*}{\begin{turn}{90}Type 4\end{turn}}
&\algone{}&0.028  &  0.082  & 0.014   & 0.003  &  0.027  & 0.097 &  0.0507 & 0.038 & 0.035 \\
&\algtwo{} & 0.232 &   0.633&   0.230&   0.265&   0.065&   0.535&   0.791&0.344 &0.279 \\
&\origekf{} &  98.424 &  198.207 &  297.957  & 97.891 & 97.811  & 97.781 &82.600&121.334 &89.050 \\
\hline
\multirow{3}{*}{\begin{turn}{90}Type 5\end{turn}}
&\algone{}     & 0.025 &   0.865   & 0.133&    0.197 &   0.318&   0.381 &  0.1123 &0.254&0.279\\
&\algtwo{}   &4.95 & 1.07  & 22.55  &  2.65   &22.02 &  1.93  & 17.63& 9.101& 9.842\\
&\origekf{}  & 162.368   &104.922  & 136.901  & 186.414& 156.470  & 149.342  & 167.329 &132.968& 58.803 
\end{tabular}
\caption{Synchronization error ($\mu$ seconds) of different nodes with respect to node 0. Type 4 attack considers adding constant values of order $\sim100 \mu$ seconds to measurement $d_ij$. Type 5 attack considers adding uniform random values of order $\sim100 \mu$ seconds to measurement $d_ij$. The three algorithms are running in parallel on the same set of attacks values to have a fair comparison.}
\label{tab:sync_err}
  \vspace{-0.8cm}
\end{table*}

\subsection{Case Study: Mobile Nodes} \label{sec:mobile}

We have shown that our proposed algorithms can be used to localize and synchronize a network of static nodes under three types of attacks. We now present the case of a heterogeneous network containing both static
and mobile nodes. We add one mobile node in the form of a CrazyFlie quadrotor to the eight anchor nodes topology used in Section \ref{sec:static}, as shown in Figure \ref{fig:full_sys}. We analyzed the results of running \algone{}, \algtwo{} and \origekf{} to localize the mobile node flying in our lab. Again, all the links are under attacks. For space limit, we will show the results under attack 1 scenario only.

\begin{figure}[tb]
\centering
\includegraphics[width = 0.42\textwidth]{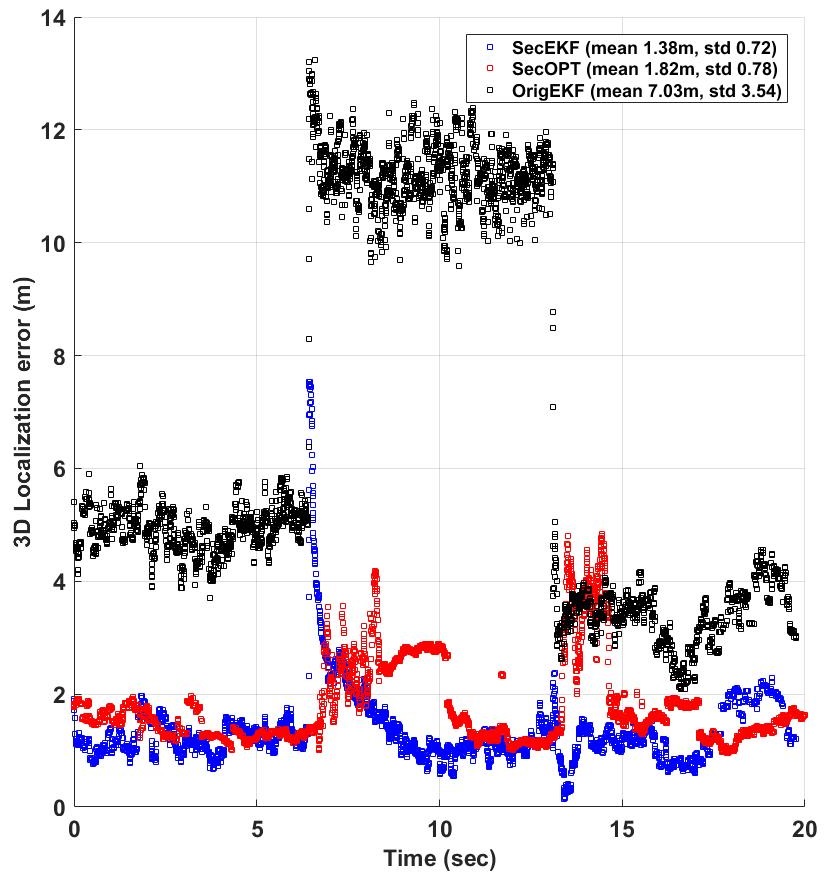}
\caption{The localization error of the  CrazyFlie quadrotor where all the measurements are under Type 1 attack with different values in a \textbf{Fully connected network}. Function \textbf{getDistanceAttackValue} shows the attacker behavior for corrupting the measurements $R_ij$ and $r_ij$. The three algorithms are running in parallel on the same set of attacks values to have a fair comparison.}
\label{fig:mobilefullyconn}
%\vspace{-0.5cm}
\end{figure}

Some experiments were performed with quadrotors traveling with variable velocities. Figure \ref{fig:mobilefullyconn} show the results of  localizing the traveling quadrotors using \algone{}, \algtwo{} and \origekf{} algorithms. \algone{} achieved the best secure localization estimation with a mean of $1.38$ m and standard deviation of $0.72$ m. On the other hand, \algtwo{} reported $1.82$m mean error and standard deviation of $0.78$ m. If the normal estimation does not consider attacks estimation, it will suffer from $7.03$m mean error and $3.54$m standard deviation.

\begin{figure}[tb]
\centering
\includegraphics[width = 0.42\textwidth]{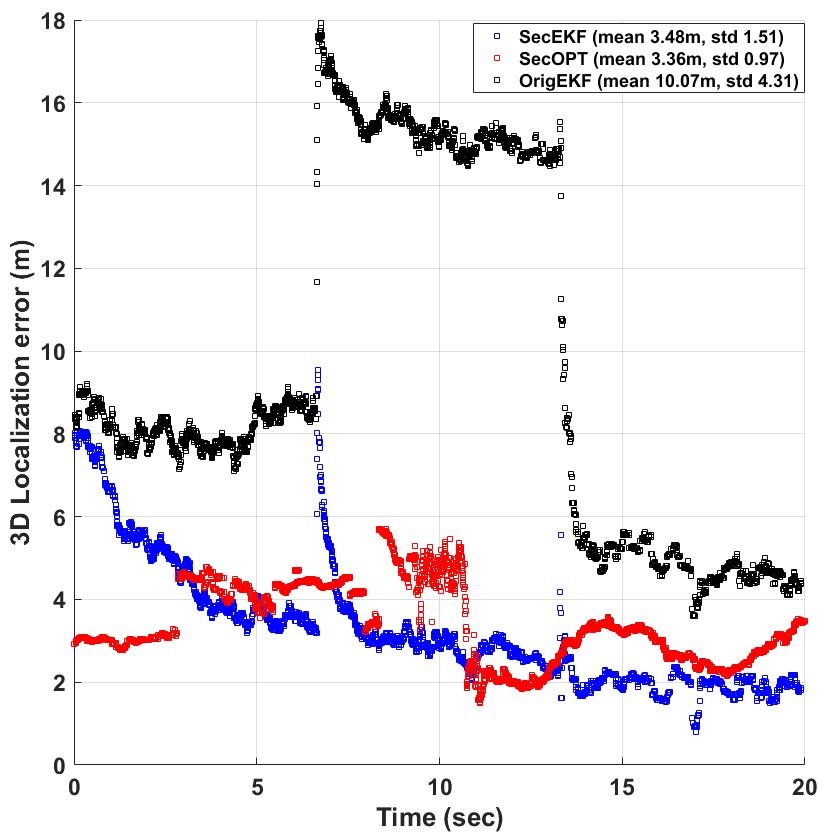}
\caption{The localization error of the  CrazyFlie quadrotor where all the measurements are under Type 1 attack with different values in a \textbf{partial connected network} where every node has five neighbors only. Function \textbf{getDistanceAttackValue} shows the attacker behavior for corrupting the measurements $R_ij$ and $r_ij$. The three algorithms are running in parallel on the same set of attacks values to have a fair comparison.}
\label{fig:mobilepartialconn}
\vspace{-0.5cm}
\end{figure}

We repeated the experiments for the network topology where every node is only connected to four neighbors instead of eight nodes in the fully connected scenario, which means we are dropping about $27$ links out of $72$ links. The performance degraded as expected. As shown in  Fig~\ref{fig:mobilepartialconn}, \algone{}, \algtwo{} and \origekf{} algorithms reported $3.48$m, $3.36$m, and $10.07$m mean localization error, respectively. 
%The results are shown in Fig~\ref{fig:mobilepartialconn}.

\subsection{Case Study: Execution Time} \label{sec:time}
\begin{figure}[tb]
\centering
\includegraphics[width = 0.43\textwidth]{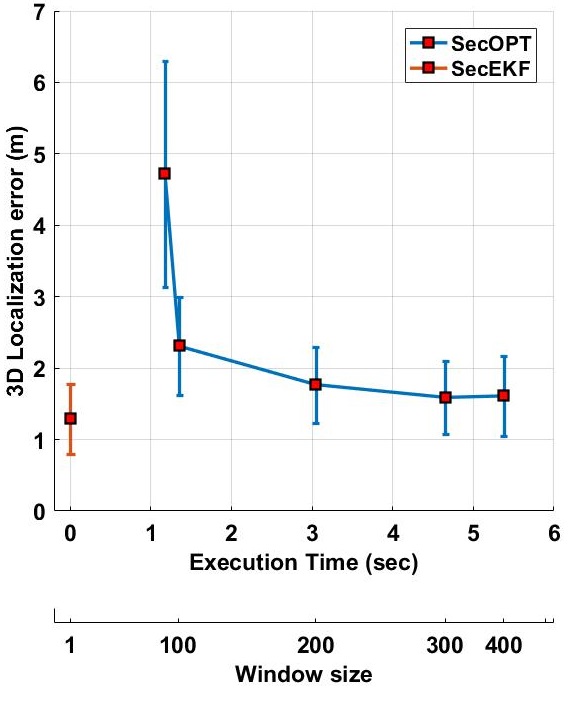}
\caption{The execution time and its corresponding localization error of the  CrazyFlie quadrotor with different window size of measurements, where all the measurements are under Type 1 attack with different values in a \textbf{Fully connected network}. Function \textbf{getDistanceAttackValue} shows the attacker behavior for corrupting the measurements $R_ij$ and $r_ij$.}
\label{fig:Exectime}
\vspace{-0.5cm}
\end{figure}

Previous work \cite{conf:attackloc} could achieve secure state estimation in case of localization in about $11$ minutes. They only consider 2D localization and only four links under attack. Eleven minutes overhead is a very high one especially for the moving nodes. We are going during this case study to analyze the required execution time for the proposed secure estimation algorithms to localize the CrazyFlie quadrotor, where all the links are under Type 1 attack. The execution of \algtwo{} depends on the number of considered measurements during one run of the algorithms which we called the window size. Figure \ref{fig:Exectime} summarizes the 3D localization error for each window size. For a window size of $300$ measurements, we could achieve a mean error of $1.59$m at the cost of $4.6$ seconds of execution time. When we decreased the window size to $50$ measurements, we get $4.7$m localization error at the cost of $1.1$ seconds of execution time. On the other hand, we can see the \algone{} could achieve a mean error of $1.28$m with a cost of $0.004$ seconds. \algone{} considers only a window size of one measurement as it depends on EKF.

\section{Conclusion} \label{sec:conc}

This work has investigated the security aware-aspect of the estimation algorithms for the multi-sensor system. More specifically, we proposed \algone{} and \algtwo{} secure estimation algorithms for the wireless network where state evolution and measurement models can follow nonlinear functions. Then, we picked simultaneous localization and synchronization as a representative application for our secure estimation problem. Several experiments in a network with static nodes and a mobile quadrotor, all equipped with commercial ultra-wideband wireless radios, were conducted.
%using real, custom ultra-wideband wireless anchor nodes and mobile quadrotor nodes were conducted. 
Our empirical results indicate that \sys{} offers reliable performance and efficient computational procedures for the secure state estimation. Our proposed algorithm made it possible by state-of-the-art advances in commercial ultra-wideband radios to securely estimate the state in asynchronous networks. Future directions will deal with testing over a large-scale system and more sophisticated attack scenarios.

%\section*{Acknowledgments}

%This research is funded in part by the National Science Foundation under awards CNS-1329755 and CNS-1329644. The U.S. Government is authorized to reproduce and distribute reprints for Governmental purposes notwithstanding any copyright notation thereon. The views and conclusions contained herein are those of the authors and should not be interpreted as necessarily representing the official policies or endorsements, either expressed or implied, of NSF, or the U.S. Government.
%\section*{Acknowledgments}
%Acknowledgement goes here.

%\bibliographystyle{abbrv}

\bibliographystyle{IEEEtran}
%{\small
\bibliography{IEEEabrv,ref}
%}
%\vspace{20cm}
%\newpage
%\clearpage
\section*{Appendix} \label{sec:appendix}

\begin{table}[t]
\centering
\normalsize
\begin{tabular}{c}
%\hline
\begin{samplelisting}[colback=white]{Attack Generator Pseudocode}
float getDistanceAttackValue(type, currentT, totalT)
{
    if(currentT < totalT/3)
    {
        Shift = 2;
        paretoScale = 3;
    }
    else if(currentT>totalT/3 && currentT<2*totalT/3)
    {   
        Shift = 6;
        paretoScale = 6.5;
    }
    else
    {
        Shift = 1;
        paretoScale = 2;
    }
    paretoShape = 3; 
    if(type==1)      
        // Uniform Distribution
        attack = 2*(rand() + Shift); 
    else if(type==2) 
        // Normal Distribution
        attack = 2*randn() + Shift;
    else if(type==3) 
        // Pareto Distribution
        attack = randp(paretoShape,paretoScale);
    return attack;
}
\end{samplelisting} 
\end{tabular} 
%\vspace{-0.6cm}
\end{table}

\subsection{Experimental Setup}
The main components of our testbed can be summarized in the following points:
\begin{figure*}[t]
\centering
\includegraphics*[width =\textwidth]{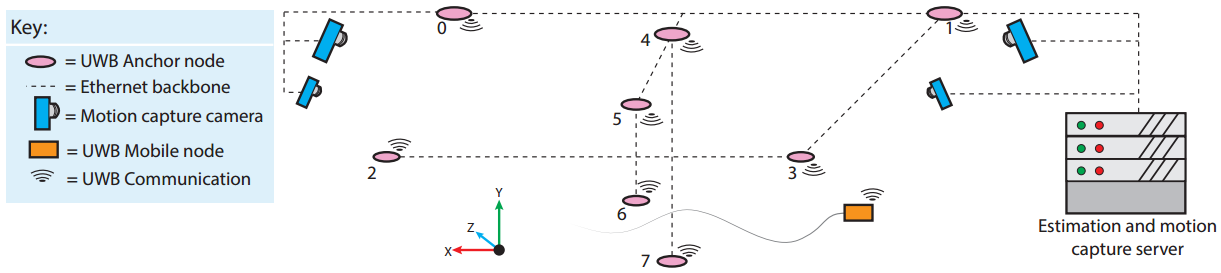}
\caption{Experimental setup overview, including, UWB nodes, motion capture cameras, and mobile quadrotor UWB nodes}
\label{fig:full_sys}
%  \vspace{-0.6cm}
\end{figure*}

\begin{figure*}
\centering
\begin{minipage}[t]{0.30\textwidth}%
\includegraphics[height=1.75in]{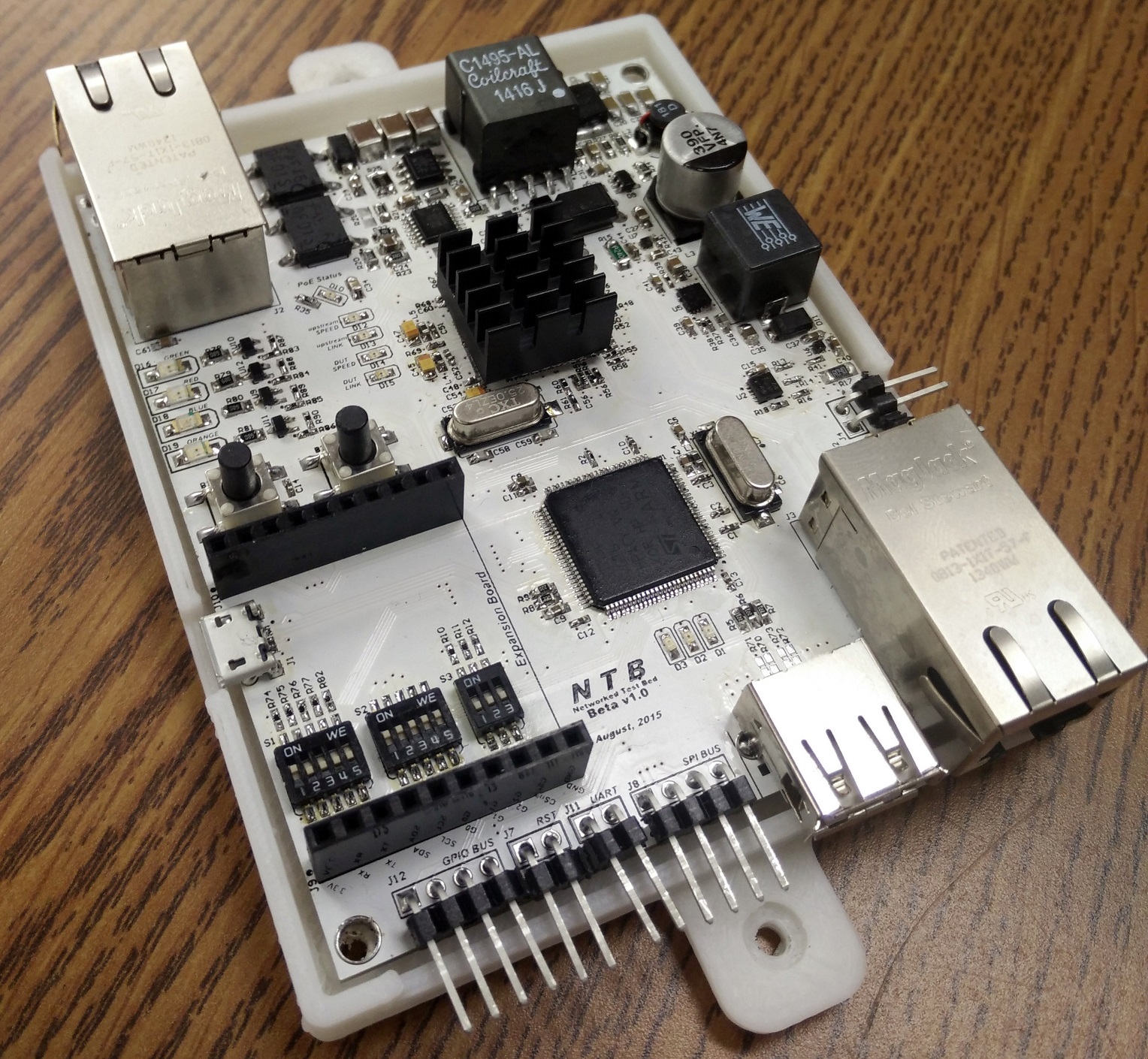}
\caption{Custom anchor node with ARM Cortex M4 processor and UWB expansion.}
\label{fig:ntb_anon}
\end{minipage}\hspace{5mm}
\begin{minipage}[t]{0.30\textwidth}%
\centering
\includegraphics[height=1.75in]{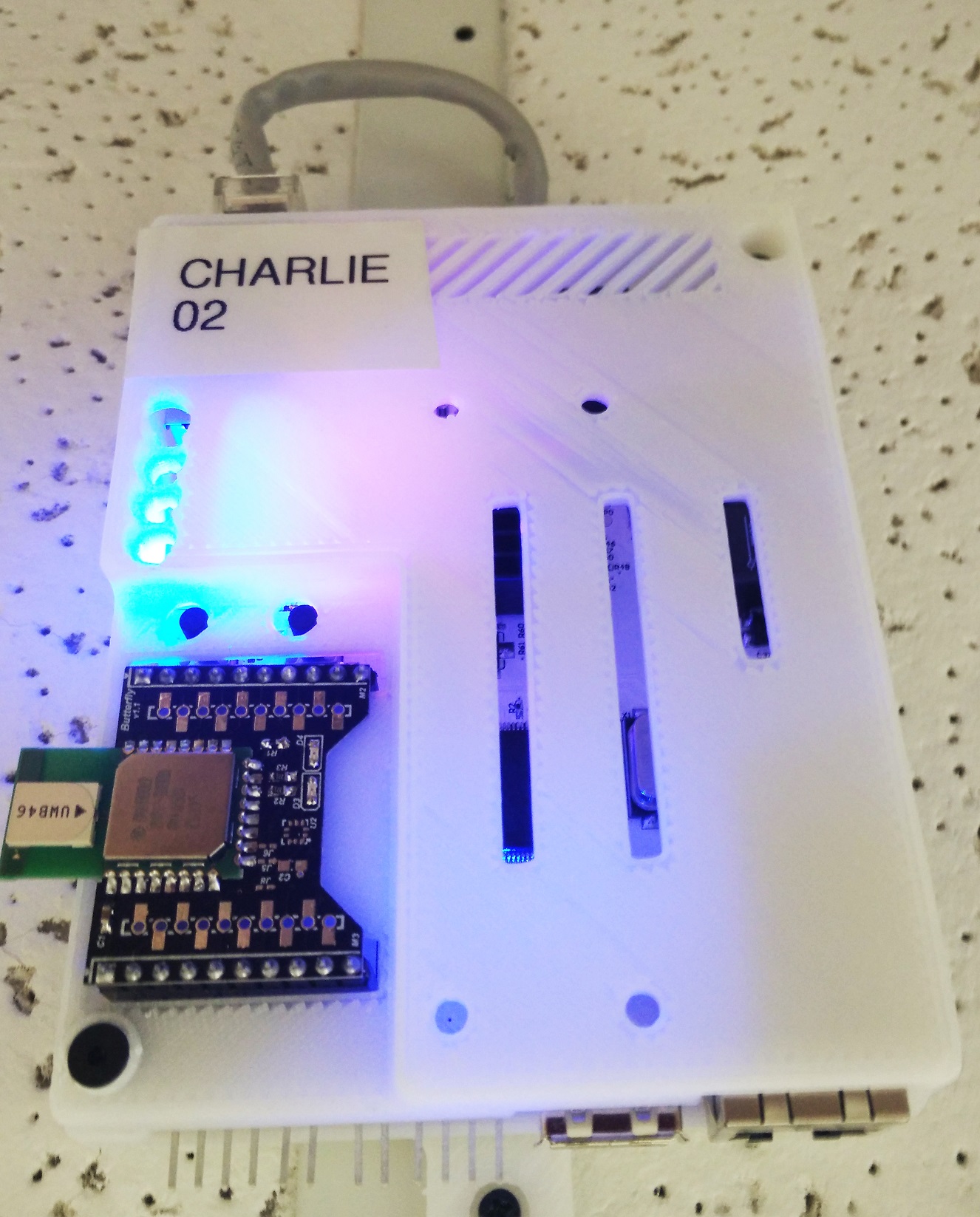}
\caption{Ceiling-mounted anchor with DW1000 UWB radio in 3D-printed enclosure.}
\label{fig:ntb_ceil}
\end{minipage}\hspace{5mm}
\begin{minipage}[t]{0.30\textwidth}%
\centering
\includegraphics[height=1.75in]{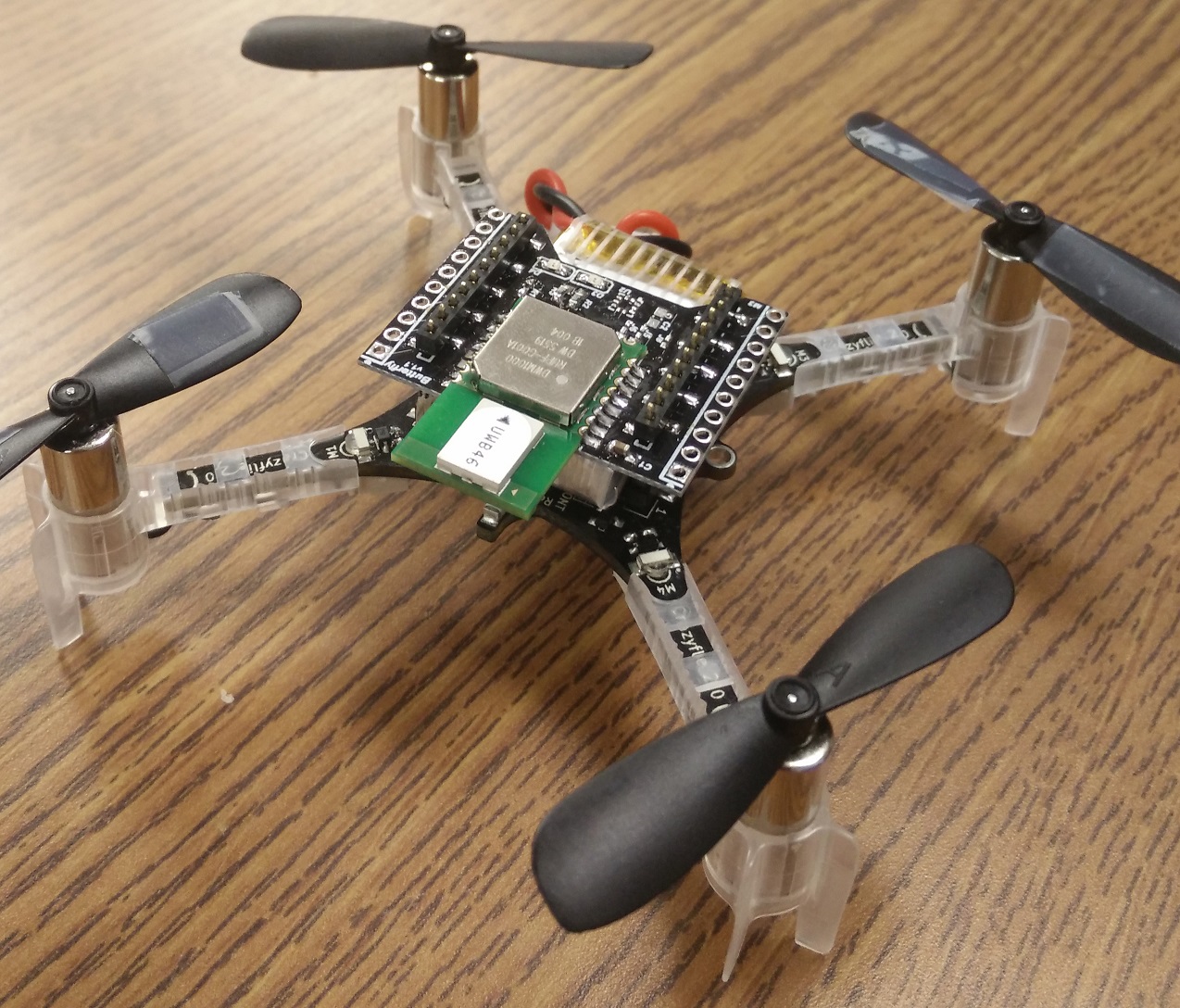}
\caption{CrazyFlie 2.0 quadrotor helicopter with DW1000 UWB expansion.}
\label{fig:quad}
\end{minipage}
\end{figure*}

\begin{itemize}

\item Motion capture system capable of 3D rigid body position measurement with less than 0.5 mm accuracy. The system consists of an eight-camera which are deployed in order to provide accurate ground truth position measurements. The results presented in this work treat the motion capture estimate as a true position, though we qualify here that all results are accurate to within the motion capture accuracy. The ground truth position estimates from the motion capture cameras are sent to a centralized server which uses the Robot Operating System (ROS) \cite{ros} with a custom package. We adopt a right-handed coordinate system where $y$ is the vertical axis, and $x$ and $z$ make up the horizontal plane. 

\item The Fixed nodes used in the following experiments consist of custom-built circuit boards equipped with ARM Cortex M4 processors t 196 MHz powered over Ethernet and communicating to Decawave DW1000 ultra-wideband radios as shown in Figure \ref{fig:ntb_anon} and Figure \ref{fig:ntb_ceil}. Each node performs a single and double-sided two-way range with its neighbors. The used Decawave radio is equipped with a temperature-compensated crystal oscillator with frequency equals $38.4$ MHz and a stated frequency stability of $\pm 2$ ppm. We installed eight UWB nodes in different positions in a 10$\times$9 $m^2$ lab. More specifically, six nodes are placed on the ceiling at about $2.5 m$ high, and two were placed at waist height at about $1$ m in order to better disambiguate positions on the vertical axis. Each node is fully controllable over a TCP/IP command structure from the central server. These nodes are placed so as to remain mostly free from obstructions, maximizing line-of-sight barring pedestrian interference. 

\item The Mobile Node used in the experiments are battery-powered mobile nodes also with ARM Cortex M4 processors based on the CrazyFlie 2.0 helicopter \cite{crazyflie} and equipped with the very same DW1000 radio as shown in Figure \ref{fig:quad}. This allows for compatibility in the single and double sided ranging technique used. 

\item The server is a MacBook Pro laptop with Intel(R) Core i7-4870HQ CPUs @2:5GHz, which is a normal PC.
\end{itemize}

\subsection{Attacks on Distances}
 Function \textbf{getDistanceAttackValue} shows the attacker behavior for corrupting the measurements. We considered, as a proof of concept, three types of attacks using Uniform, Normal and Pareto Distributions. We gave them names as type 1, 2, and 3 attacks, respectively. rand(), randn() and randp() are considered APIs to get the required distribution. currentT is the current system time and totalT is the total execution time. We considered Pareto Distribution as an example of heavy-tailed distributions. The Pareto distribution is characterized by a scale parameter $paretoScale$ and a shape parameter $paretoShape$. We picked the parameters of the distributions to be logical attack values resulting in an acceptable position in the localization area. Otherwise, the attack would be easily detected. 
 
 Note that generating the attack values according to Function \textbf{1} simulates time-varying attack values that are different for all links.
The trend in the evaluation of the related work is to use a uniform random number generator to define attack which can look like some bias in the sensor reading from a practical point of view. Thus, we used such time-varying parameters that can evaluate the algorithm in a better way.

\subsection{Attacks on Time}
We would like to open a discussion in the types of the time attacks, as they are tricky ones. Simple monitoring for evolving time of every node can help in detecting many attacks. For example, the time should not go backward; thus any reported time value in the past can be flagged as an attack. Also, the time can't go faster than the specification of the crystal, which is public information. However, we did not consider this type of smart checkers in \sys{}, we just let the secure algorithms deal with that, but it can be a possible future work. One popular attack in the time domain to provide a constant delay to pretend, as if the time is shifted, that is why we considered adding constant time attacks to the measurements while evaluating \sys{}.

\subsection{Measurements Correlations}

Many sensors in CPS reported correlated measurements. For instance, some distance measurements could be based on the time of flight calculations which is tightly coupled with time measurements. Such tight coupling could be investigated to detect attacks. We should note that \sys{} does not at this stage consider the measurements correlation and dealt with the measurements as uncorrelated measurements. Many other practical considerations could be investigated to enhance the secure estimation algorithms.

\subsection{CPS Physics}

Understanding the underlying physics governing the CPS itself would offer significant support to the secure estimation algorithms using preprocessing phase which would help in filtering the measurements. If a sensor observes
a signal that appears to violate the physics governing the
CPS dynamics, an attack flag can be easily raised to enhance the CPS security.
%\bibliographystyle{IEEEtran}
%\bibliography{ref}  % sigproc.bib is the name of the Bibliography in this case
% ACM needs 'a single self-contained file'!
%\balancecolumns
%\appendix
%Appendix A

%Appendix goes here.

% That's all folks!
\end{document}